\documentstyle[12pt]{article}
\begin{document}

\title{\hspace{11cm}{\small \bf IMPNWU-970401}\\
\vspace{2cm}\bf\huge
The elliptic quantum algebra $A_{q,p}(\widehat {sl_n})$ and
its bosonization at level one
}

\author{
{\bf
Heng Fan$^{b}$, Bo-yu Hou$^b$, Kang-jie Shi$^{a,b}$, Wen-li Yang$^{a,b}$
}\\
\normalsize $^a$ CCAST(World Laboratory)\\
\normalsize P.O.Box 8730,Beijing 100080,China\\
\normalsize $^b$ Institute of Modern Physics, P.O.Box 105,\\
\normalsize Northwest University, Xian 710069,China\\
}

\maketitle

\begin{abstract}
We extend the work of Foda et al and propose an elliptic quantum 
algebra $A_{q,p}(\widehat {sl_n})$. Similar to the case of 
$A_{q,p}(\widehat {sl_2})$, our presentation of the algebra is 
based on the relation $RLL=LLR^*$, where $R$ and $R^*$ are
$Z_n$ symmetric R-matrices with the elliptic moduli 
chosen differently and a scalar factor is also involved. With the help
of the results obtained by Asai et al, 
we realize type I and type II vertex operators
in terms of bosonic free fields for $Z_n$ symmetric Belavin model. 
We also give a bosonization for the elliptic quantum algebra
$A_{q,p}(\widehat {sl_n})$ at level one.

\end{abstract}

\section{Introduction}

The investigation of the symmetries in quantum integrable models
has been attracting a great deal
of interests. Recently, the quantum
affine algebra $U_q(\widehat {sl_2})$ 
has been studied extensively and applied successfully to the 
$XXZ$ model in the anti-ferromagnetic regime, see [1] and
the references therein. The R-matrix associated with the $XXZ$ model
is the six-vertex model which is a trigonometric vertex model.
Using the approach of free boson realization of the vertex operators,
Jimbo et al obtained integral formulae for the correlation functions
and the form factors for the $XXZ$ model.

It is well known that we can obtain the six-vertex model from
the Baxter's [2] eight-vertex model by taking a special limit.
And we know that the $XYZ$ model, spin chain equivalent of the 
eight-vertex model, is a generalization of $XXZ$ model. Foda et al [3]
proposed an elliptic extention of the quantum affine algebra
$A_{q,p}(\widehat {sl_2})$ as an algebra of symmetries for the
eight-vertex model. The Kyoto group also conjecture that type I and type
II vertex operators for the elliptic algebra $A_{q,p}(\widehat {sl_2})$
can be found. So, it is an open problem to give a bosonic free fields 
realization for vertex operators of 
the elliptic algebra $A_{q,p}(\widehat {sl_2})$.
It is also an interesting problem to enxtend the elliptic algebra
$A_{q,p}(\widehat {sl_2})$ to a more general case 
$A_{q,p}(\widehat {sl_n})$ which would play the role of the symmetry 
algebra in $Z_n$ Belavin model[11]. In this paper, we will study these 
problems.

It is now believed that the vertex operator approach [1,4] and the 
method of bosonization are very powerful to study correlation functions 
of solvable lattice models. It is firstly formulated for vertex models,
and then extended to incorporate face models[5,6]. 
Lukyanov and Pugai [6] give successfully a bosonic realization of
the vertex operators for ABF model [7]. This result is developed to 
to a more general case 
by Asai et al [8]. They give a bosonization of vertex operators for
the $A_{n-1}^{(1)}$ face model [9], the ABF model being the case $n=2$ with 
restricted condition.
We know there is a face-vertex correspondence between 
$A_{n-1}^{(1)}$ face model
and $Z_n$ Belavin [10] vertex model. When $n=2$, $Z_n$ symmetric 
Belavin model reduces to the Baxter's eight-vertex model. 
In our former work [11], 
we use the intertwiners of face-vertex correspondence, and obtained 
type I vertex operators for $Z_n$ Belavin vertex model with the help of the
results obtained by Asai et al [8]. 
The correlation functions of $Z_n$ Belavin model are
also obtained. In the present paper, continuing to extend the 
results in Ref.[8] and  our former work in Ref.[11], 
we give a bosonization of type II vertex operators for 
$Z_n$ Belavin model. Using the Miki's [12] construction, 
we obtain a bosonic realization for the elliptic algebra 
$A_{q, p}(\widehat {sl_n})$ (which is first proposed in this paper) 
at level one . 
As a special case of n=2 ,this will give the bosonization for 
$A_{q,p}(\widehat {sl_2})$ algebra at level one.

The paper is organized as follows. In Section 2 we introduce the
$Z_n$ symmetric Belavin vertex model and give a definition for
the elliptic algebra $A_{q, p}(\widehat {sl_n})$. In Section 3 we
define the vertex operators for the algebra $A_{q, p}(\widehat {sl_n})$
at level one. Section 4 is devoted to the main results of this paper.
Extending the work of Asai et al, we introduce another set of boson
oscillators and obtain a bosonization of type II vertex operators for
$A_{n-1}^{(1)}$ face model. Using face-vertex correspondence, we give
a free boson realization of type I and type II vertex operators 
for $Z_n$ Belavin model. This gives a bosonization of the elliptic
algebra $A_{q, p}(\widehat {sl_n})$ at level one. Finally, we give summary 
and discussions in Section 5. Appendix contains some detailed 
calculations.

\section{The model and the elliptic algebra $A_{q, p}(\widehat {sl_n})$}

We first introduce some notations.
Let $n\in Z^+$ and $n\ge 2$, $w\in C$ and Im$w\ge 0$, 
$r\ge n+2$ and take real value, $\tau \in C$ with Im$\tau >0$.

Define matrix $g, h, I_{\alpha }$ with elements take values
\begin{eqnarray}
g_{ij}&=&\omega ^i\delta _{ij},~~\omega =exp\{ {{2\pi i}\over n}\},  
\nonumber \\
h_{jk}&=&\delta _{j+1, k}, 
\nonumber \\
I_{\alpha }&=&I_{(\alpha _1, \alpha _2)}=g^{\alpha _2}h^{\alpha _1}.
\nonumber 
\end{eqnarray}
The elliptic functions defined as
\begin{eqnarray}
\theta \left[ \begin{array}{c}a \\b\end{array}
\right] (z, \tau )
&=&\sum _{m\in Z}exp\{\pi i[ (m+a)^2\tau +2(m+a)(z+b)]\}, 
\nonumber \\
\sigma _{\alpha }(z, \tau )&=&\sigma _{(\alpha _1, \alpha _2)}(z, \tau )
=\theta \left[ \begin{array}{c}{\frac {1}{2}+\frac {\alpha _1}{n}}\\
{\frac {1}{2}+\frac {\alpha _2}{n}}\end{array}
\right] (z, \tau ).
\nonumber 
\end{eqnarray}
Redefine $z\equiv vw, \tau\equiv rw $. The R-matrix of $Z_n$ symmetric 
R-matrix can be defined as
\begin{eqnarray}
\bar {R}(vw, rw)=
\frac {\sigma _0(w, rw)}{\sigma _0(vw+w, rw)}
\sum _{\alpha }W_{\alpha }(vw, rw)I_{\alpha }\otimes I_{\alpha }^{-1},
\end{eqnarray}
where 
\begin{eqnarray}
W_{\alpha }(wv, rw)=
\frac {\sigma _{\alpha }(vw+{w\over n}, rw)}
{n\sigma _{\alpha }({w\over n}, rw)}.
\end{eqnarray}
It is necessary to introduce other notations
\begin{eqnarray}
R(v)\equiv R(vw, rw)=x^{2v({1\over n}-1)}
\frac {g_1(v)}{g_1(-v)}
\bar {R}(vw, rw),
\end{eqnarray}
where $x=e^{\pi iw}$ and 
\begin{eqnarray}
g_1(v)=\frac {\{ x^{2v}x^2\} \{ x^{2n+2r-2}x^{2v}\} }
{\{ x^{2n}x^{2v}\} \{ x^{2r}x^{2v}\} }
\end{eqnarray}
where $\{ z\} =(z; x^{2r}, x^{2n})$ and 
$(z; p_1, \cdots ,p_n)\equiv \prod _{\{n_i\}=0}^\infty 
(1-zp_1^{n_1}\cdots p_m^{n_m})$.

The R-matrix $R(v, rw)$ have the following properties:
\begin{eqnarray}
{\rm Yang-Baxter\ \ equation}&:&R_{12}(v_{1}-v_{2})R_{13}(v_{1}-v_{3})
R_{23}(v_{2}-v_{3})\nonumber\\
& &\ \ \ \ \ =R_{23}(v_{2}-v_{3})R_{13}(v_{1}-v_{3})
R_{12}(v_{1}-v_{2})\nonumber\\
{\rm Unitarity}&:& R_{12}(v)R_{21}(-v)=1,
\nonumber \\
{\rm Cross-unitarity}&:& \sum _{jl}R_{kj}^{li}(v)
R_{i'l}^{jk'}(-v-n)=\delta _{ii'}\delta _{kk'}.
\end{eqnarray}
The parameters $v$,$w$,$r$ in our parameterization for $Z_n$ symmetric 
R-matrix in Eq.(3) are related to that of Foda et al [8] as follows:
$q=e^{i\pi w}=x$ , $\xi =x^{v}$ , $p=x^{2r}$ .

Define
\begin{eqnarray}
R^*(v)=\Delta _n^2(v)R(vw, (r-c)w),
\nonumber \\
\Delta _n(v)=-x^{\frac {2(n-2)}{n}v}
\frac {(x^{2n-2+2v};x^{2n})(x^{2-2v};x^{2n})}
{(x^{2+2v};x^{2n})(x^{2n-2-2v};x^{2n})}
\end{eqnarray}
{\bf Definition}: {\it Algebra $A_{q,p}(\widehat {gl_n})$ is generated
by $L_{ij}(v)$ satisfying the relation}
\begin{eqnarray}
& &R^+(v_1-v_2)L_1(v_1)L_2(v_2)
=L_2(v_2)L_1(v_1)R^{*+}(v_1-v_2),
\end{eqnarray}
{\it where
\begin{eqnarray}
& &R^+(v)=R(v)\tau  ^{-1}(-v+{1\over 2})\ \ ,\ \ 
R^{*+}(v)=R^*(v)\tau ^{-1}(-v+{1\over 2})\nonumber\\
& &\tau (v)=x^{\frac {2(1-n)}{n}v}
\frac {(x^{1+2v};x^{2n})(x^{2n-1-2v};x^{2n})}
{(x^{2n-1+2v};x^{2n})(x^{1-2v};x^{2n})}
\end{eqnarray}}

By a standard argument based on the anti-symmetric fusion for $Z_n$ symmetry 
R-matrix [10,15],we find that the following quantum determinant belongs to 
the center of $A_{q,p}(\widehat {gl_n})$:
\begin{eqnarray}
q-det\ \ L(v)=\sum_{\sigma \in S_n}sign(\sigma )L_{1,\sigma (1)}(v-n)
L_{2,\sigma (2)}(v-n+1)....L_{n,\sigma (n)}(v-1)
\end{eqnarray}
Therefore , we can impose the further relation 
$q-det\ \ L(v)=q^{\frac{c}{2}}$ and define the quotient algebra 
\begin{eqnarray}
A_{q,p}(\widehat {sl_n})=A_{q,p}(\widehat {gl_n})/
(q-det\ \ L(v)-q^{\frac{c}{2}})
\end{eqnarray}

{\bf Remark}:{\it For the case $n=2$, we have $\Delta _n(v)=-1$. The
algebra $A_{q,p}(\widehat {sl_n})$ reduces to the original elliptic
algebra $A_{q,p}(\widehat {sl_2})$ proposed by Foda et al[3].}

\section{Bosonization of the algebra 
$A_{q,p}(\widehat {sl_n})$ at level one}
In the following, we will mainly restrict our attention to the level one case,
and we have $c=1$.The $R^{*}(v)$ now becomes  
\begin{eqnarray}
R^*(v)=\Delta _n^2(v)R(v, (r-1)w).
\end{eqnarray}
The algebra relation remains as:
\begin{eqnarray}
R^+(v_1-v_2)L_1(v_1)L_2(v_2)
=L_2(v_2)L_1(v_1)R^{*+}(v_1-v_2).
\end{eqnarray}

We first introduce type I vertex operator corresponding to the half-column 
transfer 
matrix of $Z_n$ Belavin model [5,11,13],and type II vertex operator of
$Z_n$ Belavin model which are expected to create the eigenstates of the 
transfer matrix .We denote the two types of vertex operators as: 
\begin{itemize}
\item Vertex operator of type I: $\Phi _i(v)$,
\item Vertex operator of type II: $\Psi ^*_i(v).$
\end{itemize}
These Vertex operators satisfy the Faddeev-Zamolodchikov (ZF) algebra.                           
\begin{eqnarray}
\Phi _j(v_2)\Phi _i(v_1)&=&\sum_{lk}R_{lk}^{ij}(v_1-v_2)
\Phi _l(v_1)\Phi _k(v_2),
\\
\Psi _i^*(v_1)\Psi _j^*(v_2)&=&\sum_{lk}\Psi _k^*(v_2)\Psi _l^*(v_1)
{R^*}^{lk}_{ij}(v_1-v_2)\Delta _n^{-1}(v_1-v_2),
\\
\Phi _i(v_1)\Psi _j^*(v_2)&=&\tau ^{-1}(v_1-v_2)
\Psi _j^*(v_2)\Phi _i(v_1).
\end{eqnarray}
In the next section, we will give a q-boson free fields realization
of the type I and type II vertex operators listed above.

Introduce Miki's construction [12],
\begin{eqnarray}
L_{ij}(v)=\Phi _i(v)\Psi _j^*(v-{1\over 2}).
\end{eqnarray}
Using relations of the ZF algebra Eq.(13)---Eq.(15), one can prove that
the operator matrix $L$ constructed above satisfy the defining relation
of the elliptic quantum algebra Eq.(7). Here the relation 
$\frac {\tau (v+{1\over 2})}
{\tau (-v+{1\over 2})}=
\Delta _n^{-1}(v)$ has been used.

Thus if the bosonization of the type I and type II vertex
operators at level one satisfying Eq.(13)---Eq.(15) can be find, we can find 
a  free boson realization of the elliptic algebra $A_{q,p}(\widehat {sl_n})$ 
at level one.

\section{Bosonization for vertex operators}
In this section, we will use face-vertex correspondence relation 
to obtain a bosonization of vertex operators for $Z_n$ Belavin model 
satisfying  Eq.(13)---Eq.(15).

\subsection{Bosonization for $A_{n-1}^{(1)}$ face model}
We first give a brief review of the 
bosonization of type I vertex operators for $A^{(1)}_{n-1}$ face model 
[8],and then similarly constructe the bosonization of type II vertex 
operators for $A^{(1)}_{n-1}$ face model. 

Let $\epsilon_{\mu}$ $(1\le \mu \le n)$ be the orthonormal basis in $R^{n}$ ,  
which are supplied with the inner product $<\epsilon_{\mu},\epsilon_{\nu}>=
\delta_{\mu \nu}$ .Set 
\begin{eqnarray*}
\overline{\epsilon}_{\mu}=\epsilon_{\mu}-\epsilon\ \ ,\ \ \epsilon=
\frac{1}{n}\sum_{\mu =1}^{n}\epsilon_{\mu}.
\end{eqnarray*}
The type $A^{(1)}_{n-1}$ weight Lattice is the linear span of the 
$\overline{\epsilon}_{\mu}$ : $P=\sum^{n}_{\mu =1}Z
\overline{\epsilon}_{\mu}$.
Let $\omega_{\mu} (1\le \mu \le n-1)$ be the fundmental weights and 
$\alpha_{\mu} (1\le \mu \le n-1)$ be the simple roots: 
$\omega_{\mu}=\sum_{\nu =1}^{\mu}\overline{\epsilon}_{\nu}\ \ ,\ \ 
\alpha_{\mu}=\epsilon_{\mu}-\epsilon_{\mu +1}$ .

An ordered pair $(b,a) \in P^{2}$ is called admissible if and only if there 
exists $\mu (1\le \mu \le n)$ such that $ b-a=\overline{\epsilon}_{\mu}$. 
An ordered set of four weights $\left( \begin{array}{ll}
c&d\\b&a\end{array}\right)\in P^{4}$ is called an admissible configuration
around a face if and only if the pairs (b,a),(c,b),(d,a) and (c,d) are 
admissible. To each admissible configuration around a face , one can 
associate the Botlzmann weight for it [2,7,9].  

We introduce the zero mode operators $q_{\mu}$ , $p_{\mu}$ $(1\le \mu \le n)$ ,
 which satisfy:
\begin{eqnarray*}
[p_{\mu},iq_{\nu}]=<\epsilon_{\mu},\epsilon_{\nu}>=
\delta_{\mu\nu} .
\end{eqnarray*}
\begin{eqnarray*}
{\rm Set }\ \ \ \ \ \ \ \  
Q_{\overline{\epsilon}_{\mu}}=q_{\mu}-\frac{1}{n}\sum_{l=1}^{n}q_{l}\ \ ,\ \ 
P_{\overline{\epsilon}_{\mu}}=p_{\mu}-\frac{1}{n}\sum_{l=1}^{n}p_{l}
+\frac{1}{\sqrt{r(r-1)}}w_{\mu} ,
\end{eqnarray*}
\noindent where $\{w_{j}\}$ are  generic complex numbers,which ensures that 
the intertwiners $\tilde {\varphi} (v) ^{(k)}_{\mu,a}$ and 
$\bar {\varphi}(v)^{(k)}_{a,\mu}$ (see the following subsection)  
could exist . We can also  reconstruct the zero mode operators 
$P_{\alpha}$ , $Q_{\alpha}$ [8] indexed by $\alpha \in P$ ,which are 
Z-linear in $\alpha$ and satisfy 
\begin{eqnarray*}
[P_{\alpha},iQ_{\beta}]=<\alpha ,\beta> \ \ \ \ (\alpha ,\beta \in P).
\end{eqnarray*}
Now we consider the oscillator part.Define free bosonic oscillators 
$\beta^{j}_{m}\ \ (1\le j\le n,m\in Z/\{ 0\} )$ satisfying relations
\begin{eqnarray}
[\beta _m^j, \beta _n^k]
=\left\{ 
\begin{array}{l}
m\frac {[(n-1)m]_x[(r-1)m]_x}
{[nm]_x[rm]_x}\delta _{n+m,0}, ~~~j=k
\\
-mx^{sign(j-k)nm}
\frac {[m]_x[(r-1)m]_x}
{[nm]_x[rm]_x}\delta_{n+m,0}, ~~~j\not= k.
\end{array}
\right.
\end{eqnarray}
Here we have use the standard notation $[a]_x=\frac {x^a-x^{-a}}{x-x^{-1}}$.
The bosonic oscillators should satisfy the constraint:
$\sum _{j=1}^nx^{-2jm}\beta _m^j=0$. one can check that the above 
constraint is compatable with  the commutation relations Eq.(18).
Similarly, we can also introduce another set of bosonic oscillators
${\beta '}_m^j~(1\le j\le n, m\in Z/\{ 0\} )$ which will be used to 
constructe the type II vertex operators and are related to the original 
bosons by: ${\beta '}_m^j=
\frac {[rm]_x}{[(r-1)m]_x}\beta _m^j$.
Use $\beta _m^j, {\beta '}_m^j$, we define operators 
\begin{eqnarray}
S_m^j=(\beta _m^j-\beta _m^{j+1})x^{-jm},
~~\Omega _m^j=\sum _{k=1}^jx^{(j-2k+1)m}\beta _m^k,\\
{S'}_m^j=({\beta '}_m^j-{\beta '}_m^{j+1})x^{-jm},
~~{\Omega '}_m^j=\sum _{k=1}^jx^{(j-2k+1)m}
{\beta' }^{k}_{m}.
\end{eqnarray}
Define the bosonic Fock spaces
${\cal {F}}_{l,k}=c[\{ \beta _{-1}^j, \beta _{-2}^j, 
\cdots \}_{1\le j\le n}]|l, k>$ and the vacuum vector
\begin{eqnarray}
|l,k>=e^{i\sqrt {\frac {r}{r-1}}Q_l 
-i\sqrt {\frac {r-1}{r}}Q_k}|0,0>
\end{eqnarray}
In the following part of this paper, we will construct the bosonic 
realization  of the vertex operators defined in Eq.(13)---Eq.(15) 
on the space $\sum_{l,k\in P}\oplus F_{l,k}$. For the generic r (e.g  
irrational number) , we can introduce two type independent operators  
$\widehat{K}$ and $\widehat{L}$ with 
$\widehat{K}=\sum_{\mu}\widehat{k}_{\mu}\overline{\epsilon}_{\mu}$ , 
$\widehat{L}=\sum_{\mu}\widehat{l}_{\mu}\overline{\epsilon}_{\mu}$ , which  
act on the space $\sum_{l,k\in P}\oplus F_{l,k}$ as 
\begin{eqnarray*}
& &\widehat{k}_{\mu}F_{l,k}=(w_{\mu}+k_{\mu})F_{l,k}\ \ ,\ \ {\rm if}\ \ \ 
k=\sum_{\mu}k_{\mu}\overline{\epsilon}_{\mu}\\
& &\widehat{l}_{\mu}F_{l,k}=(w_{\mu}+l_{\mu})F_{l,k}\ \ ,\ \ {\rm if}\ \ \ 
l=\sum_{\mu}l_{\mu}\overline{\epsilon}_{\mu}
\end{eqnarray*}

Let us introduce some basic operators acting on the space 
$\sum_{l,k\in P}\oplus F_{l,k}$
\begin{eqnarray}
\eta _j(v)&=&e^{-i\sqrt {\frac {r-1}{r}}
(Q_{\omega _j}-i2vlnxP_{\omega _j})}
:e^{-\sum _{m\not= 0}{1\over m}\Omega _m^jx^{-2vm}}:
\nonumber \\
\xi _j(v)&=&e^{i\sqrt {\frac {r-1}{r}}(Q_{\alpha _j}-i2vlnxP_{\alpha _j})}
:e^{\sum _{m\not= 0}{1\over m}S _m^jx^{-2vm}}:
\nonumber \\
\eta '_j(v)&=&e^{i\sqrt {\frac {r}{r-1}}
(Q_{\omega _j}-i2vlnxP_{w_j})}
:e^{\sum _{m\not= 0}{1\over m}{\Omega '}_m^jx^{-2vm}}:
\nonumber \\
\xi '_j(v)&=&e^{-i\sqrt {\frac {r}{r-1}}(Q_{\alpha _j}-i2vlnxP_{\alpha _j})}
:e^{-\sum _{m\not= 0}{1\over m}{S'} _m^jx^{-2vm}}:
\end{eqnarray}
We can obtain the normal order relations
which will be presented in Appendix. From those results, 
we have the following commutation relations
\begin{eqnarray}
\xi _j(v_1)\xi _{j+1}(v_2)&=&-
\frac {[v_1-v_2+{1\over 2}]}
{[v_1-v_2-{1\over 2}]}\xi _{j+1}(v_2)\xi _j(v_1),
\nonumber \\
\xi _j(v_1)\eta _j(v_2)&=&-
\frac {[v_1-v_2+{1\over 2}]}
{[v_1-v_2-{1\over 2}]}\eta _j(v_2)\xi _j(v_1),
\nonumber \\
\xi _j(v_1)\xi _j(v_2)&=&
\frac {[v_1-v_2-1]}
{[v_1-v_2+1]}\xi _j(v_2)\xi _j(v_1),
\end{eqnarray}
where 
\begin{eqnarray}
[v]=x^{\frac {v^2}{r}-v}(x^{2v};x^{2r})
(x^{2r}x^{-2v};x^{2r})(x^{2r};x^{2r})=
\sigma ({v\over r}, -{1\over {rw}})\times const.
\end{eqnarray}
The bosonization of type I vertex operator in $A^{(1)}_{n-1}$ face model 
was given by Asai et al[8].
\begin{eqnarray}
\phi _{\mu }(v)
&=&\oint \prod _{j=1}^{\mu -1}
\frac {dx^{2v_j}}{2\pi ix^{2v_j}}
\eta _1(v)\xi _1(v_1)\cdots \xi _{\mu -1}(v_{\mu -1})
\prod _{j=1}^{\mu -1}f(v_j-v_{j-1}, \pi _{j\mu }),
\\
\pi _{\mu }&=&\sqrt {r(r-1)}P_{\bar {\epsilon }_{\mu }},
~~\pi _{\mu \nu}=\pi _{\mu }-\pi _{\nu },~~v_0=v
\ \ ,\ \ f(v,w)=\frac {[v+{1\over 2}-w]}{[v-{1\over 2}]}.
\end{eqnarray}
The integration contours are simple closed curves around
the origin satisfying
\begin{eqnarray*}
x|x^{2v_{j-1}}|<|x^{2v_j}|<x^{-1}|x^{2v_{j-1}}|
\end{eqnarray*}
We have also another set of relations
\begin{eqnarray}
\xi '_j(v_1)\xi '_{j+1}(v_2)&=&-
\frac {[v_1-v_2-{1\over 2}]'}
{[v_1-v_2+{1\over 2}]'}\xi '_{j+1}(v_2)\xi '_j(v_1),
\nonumber \\
\xi '_j(v_1)\eta '_j(v_2)&=&-
\frac {[v_1-v_2-{1\over 2}]'}
{[v_1-v_2+{1\over 2}]'}\eta '_j(v_2)\xi '_j(v_1),
\nonumber \\
\xi '_j(v_1)\xi _j(v_2)&=&
\frac {[v_1-v_2+1]'}
{[v_1-v_2-1]'}\xi '_j(v_2)\xi '_j(v_1),
\end{eqnarray}
where

\begin{eqnarray}
[v]'&=&x^{\frac {v^2}{r-1}-v}(x^{2v};x^{2r-2})
(x^{2r-2}x^{-2v};x^{2r-2})(x^{2r-2};x^{2r-2})
\nonumber \\
&=&\sigma ({v\over {r-1}}, -{1\over {(r-1)w}})\times const'.
\end{eqnarray}

Construct type II vertex operators in $A^{(1)}_{n-1}$ face model as:
\begin{eqnarray}
\phi '_{\mu }(v)
&=&\oint \prod _{j=1}^{\mu -1}
\frac {dx^{2v_j}}{2\pi ix^{2v_j}}
\eta '_1(v)\xi '_1(v_1)\cdots \xi '_{\mu -1}(v_{\mu -1})
\prod _{j=1}^{\mu -1}f'(v_j-v_{j-1}, \pi _{j\mu }),
\\
f'(v,w)&=&\frac {[v-{1\over 2}+w]'}{[v+{1\over 2}]'}.
\end{eqnarray}
Here the integration contours are simple closed curves around
the origin satisfying
\begin{eqnarray*}
x|x^{2v_{j-1}}|<|x^{2v_j}|<x^{-1}|x^{2v_{j-1}}|
\end{eqnarray*}

Define the face Boltzmann weights for admissble configurations around a face 
\begin{eqnarray}
\bar {W}\left(\begin{array}{cc}
a+2\bar {\epsilon }_{\mu } &a+\bar {\epsilon }_{\mu }
\\
a+\bar {\epsilon }_{\mu } &a
\end{array}|v\right)
&\equiv &\bar{W}(a|v)^{\mu\mu}_{\mu\mu}=1
\nonumber \\
\bar {W}\left(\begin{array}{cc}
a+\bar {\epsilon }_{\mu }+\bar {\epsilon }_{\nu } 
&a+\bar {\epsilon }_{\nu }
\\
a+\bar {\epsilon }_{\nu } &a
\end{array}|v\right)
&\equiv &\bar{W}(a|v)^{\mu\nu}_{\nu\mu}=\frac {[v+a_{\mu \nu }][1]}
{[v+1][a_{\mu \nu }]},~~\mu \not= \nu,
\nonumber \\
\bar {W}\left(\begin{array}{cc}
a+\bar {\epsilon }_{\mu }+\bar {\epsilon }_{\nu } 
&a+\bar {\epsilon }_{\mu }
\\
a+\bar {\epsilon }_{\nu } &a
\end{array}|v\right)
&\equiv &\bar{W}(a|v)^{\mu\nu}_{\mu\nu}=\frac {[v][a_{\mu \nu }-1]}
{[v+1][a_{\mu \nu }]},~~\mu \not= \nu .
\end{eqnarray}
We  also define another Boltzmann weights 
\begin{eqnarray}
\bar {W}'\left(\begin{array}{cc}
a+2\bar {\epsilon }_{\mu } &a+\bar {\epsilon }_{\mu }
\\
a+\bar {\epsilon }_{\mu } &a
\end{array}|v\right)
&\equiv &\bar{W'}(a|v)^{\mu\mu}_{\mu\mu}= 1
\nonumber \\
\bar {W}'\left(\begin{array}{cc}
a+\bar {\epsilon }_{\mu }+\bar {\epsilon }_{\nu } 
&a+\bar {\epsilon }_{\nu }
\\
a+\bar {\epsilon }_{\nu } &a
\end{array}|v\right)
&\equiv &\bar{W'}(a|v)^{\mu\nu}_{\nu\mu}=\frac {[v+a_{\mu \nu }]'[1]'}
{[v+1]'[a_{\mu \nu }]'},~~\mu \not= \nu,
\nonumber \\
\bar {W}'\left(\begin{array}{cc}
a+\bar {\epsilon }_{\mu }+\bar {\epsilon }_{\nu } 
&a+\bar {\epsilon }_{\mu }
\\
a+\bar {\epsilon }_{\nu } &a
\end{array}|v\right)
&\equiv &\bar{W'}(a|v)^{\mu\nu}_{\mu\nu}=\frac {[v]'[a_{\mu \nu }-1]'}
{[v+1]'[a_{\mu \nu }]'},~~\mu \not= \nu .
\end{eqnarray}

Following the results obtained by Asai et al in ref.[8],we have 
\begin{eqnarray}
\phi _{\mu }(v_2)\phi _{\nu }(v_1)&=&r_1(v_2-v_1)
\sum _{\mu ',\nu '}\phi _{\mu '}(v_1)
\phi _{\nu '}(v_2)
\bar {W}(\hat{\pi}|v_{1}-v_{2})^{\mu\nu}_{\nu '\mu '}.
\end{eqnarray}
Using the method introduced by Asai et al in ref.[8] and noting that 
\begin{eqnarray*}
f'(v,\pi_{j\mu}-r)&=&\frac{[v-\frac{1}{2}+\pi_{j\mu}-r]'}{[v+\frac{1}{2}]'}
=-\frac{[v-\frac{1}{2}+\pi_{j\mu}-1]'}{[v+\frac{1}{2}]'}\\
&=&-\frac{[-v+\frac{1}{2}-\pi_{j\mu}+1]'}{[-v-\frac{1}{2}]'}
=-f(-v,\pi_{j\mu}-1)|_{r\longrightarrow r-1},\\
f'(v,\pi_{\mu\nu}+r)&=&-f(-v,\pi_{\mu\nu}+1)|_{r\longrightarrow r-1},
\end{eqnarray*}
we can derive the following commutation relations:
\begin{eqnarray}
& &\phi _{\mu }'(v_1)\phi '_{\nu }(v_2)=r'_1(v_1-v_2)
\sum _{\mu ',\nu '}\phi '_{\mu '}(v_2)
\phi '_{\nu '}(v_1)
\bar {W'}^{\mu\nu}_{\nu '\mu '}(\widehat{\pi}|v_{1}-v_{2}),\\
& &\phi _{\mu }(v_1)
\phi '_{\nu }(v_2)=\tau (v_1-v_2)
\phi '_{\nu }(v_2)\phi _{\mu }(v_1),
\end{eqnarray}
where
\begin{eqnarray*}
r_1(v)=x^{
\frac {2(r-1)(n-1)v}{nr}}
\frac {g_1(-v)}{g_1(v)}\ \ ,\ \ 
r'_1(v)=x^{
\frac {2r(n-1)v}{n(r-1)}}
\frac {g'_1(-v)}{g'_1(v)}.
\end{eqnarray*}
One can see that $\phi_{\mu}$ , $\phi'_{\mu}$ are the intertwiners 
between the Fock space  
\begin{eqnarray*}
& & \phi_{\mu}(v) \ \ \ \ :\ \ \ F_{l,k}\longrightarrow F_{l,k+
\overline{\epsilon}_{\mu}}\\
& & \phi'_{\mu}(v) \ \ \ \ :\ \ \ F_{l,k}\longrightarrow F_{l+
\overline{\epsilon}_{\mu},k}
\end{eqnarray*}

\subsection{Face-vertex correspondence and modular transformation}
It is convenient to introduce some notations.Define 
\begin{eqnarray}
& &\bar {R}^{(1)}(v)
=\frac {\sigma_{0} ({1\over r},-{1\over {rw}})}
{\sigma_{0} (\frac {v+1}{r}, -{1\over {rw}})}
\sum _{\alpha }
\frac {\sigma_{\alpha} ({v\over r}+{1\over {nr}},-{1\over {rw}})}
{n\sigma_{\alpha} ({1\over {nr}}, -{1\over {rw}})}
I_{\alpha }\otimes I^{-1}_{\alpha },
\nonumber \\
& &{\bar {R'}}^{(1)}(v)
=\frac {\sigma_{0} ({1\over {r-1}},-{1\over {(r-1)w}})}
{\sigma_{0} (\frac {v+1}{r-1}, -{1\over {(r-1)w}})}
\sum _{\alpha }
\frac {\sigma_{\alpha} ({v\over {r-1}}+{1\over {n(r-1)}},-{1\over {(r-1)w}})}
{n\sigma_{\alpha} ({1\over {n(r-1)}}, -{1\over {(r-1)w}})}
I_{\alpha }\otimes I^{-1}_{\alpha },
\nonumber \\
& &\varphi _{\mu ,\widehat{K}}^{(k)}(v)=\theta ^{(k)}
(\frac {v+n<\widehat{K},\epsilon _{\mu }>}{r}+
\frac {(n-1)}{r}, -{1\over {rw}}),
\nonumber \\
& &{\varphi '}_{\widehat{L},\mu}^{(k)}(v)=\theta ^{(k)}
(\frac {v+n<\widehat{L},\epsilon _{\mu }>}{r-1}+P_{0}, 
-{1\over {(r-1)w}}),\\
& &\theta ^{(k)}(z,\tau )=\theta \left[
\begin{array}{c}
-\frac{k}{n}\\
0 \end{array}\right] (z,n\tau ),\nonumber
\end{eqnarray}

Although we choose the different function $\theta^{(k)}(v)$ from 
the ordinary one[15] ,the face-vertex correspondence relation is 
still survived
\begin{eqnarray*}
& &\sum _{k,l}
\bar {R}^{(1)}(v_1-v_2)^{kl}_{ij}
\varphi ^{(k)}_{\nu ,\widehat{K}+\bar {\epsilon }_{\mu }}(v_1) 
\varphi ^{(l)}_{\mu ,\widehat{K}}(v_2)
=\sum _{\mu ',\nu '}
\bar {W}(\widehat{K}|v_{1}-v_{2})^{\nu\mu}_{\nu '\mu '}
\varphi ^{(i)}_{\nu ',\widehat{K}}(v_1)
\varphi ^{(j)}_{\mu ',\widehat{K}+\bar {\epsilon }_{\nu }}(v_2)
\end{eqnarray*}
\begin{eqnarray*}
& &\sum _{k,l}
\bar {R'}^{(1)}(v_1-v_2)^{kl}_{ij}
{\varphi '}^{(k)}_{\widehat{L} ,\mu }(v_1) 
{\varphi '}^{(l)}_{\widehat{L}-\bar {\epsilon }_{\mu },\nu }(v_2)
=\sum _{\mu ',\nu '}
\bar {W'}(\widehat{L}|v_{1}-v_{2})^{\mu\nu}_{\mu '\nu '}
{\varphi '}^{(i)}_{\widehat{L}-\bar {\epsilon }_{\nu '},\mu '}(v_1)
{\varphi '}^{(j)}_{\widehat{L},\nu '}(v_2)
\end{eqnarray*}
For the generic number $\{w_{j}\}$ ,we can introduce intertwiners 
$\tilde {\varphi }_{\mu ,\widehat{K}}(v)$ and
$\bar {\varphi }'_{\widehat{L},\mu }(v)$ satisfying relations[15]
\begin{eqnarray*}
\sum _k\tilde {\varphi }^{(k)}_{\mu ,\widehat{K}}(v)
\varphi ^{(k)}_{\nu ,\widehat{K}}(v)=\delta _{\mu \nu }\ \ ,
\ \ 
\sum _k\bar {\varphi '}^{(k)}_{\widehat{L},\mu }(v)
{\varphi '}^{(k)}_{\widehat{L},\nu }(v)=\delta _{\mu \nu }.
\end{eqnarray*}
So it is easy to find the following face-vertex correspondence relations
\begin{eqnarray}
& &\sum _{k,l}\bar {R}^{(1)}(v_1-v_2)^{ij}_{kl}
\tilde {\varphi }^{(k)}_{\mu ,\widehat{K}}(v_1)
\tilde {\varphi }^{(l)}_{\nu ,\widehat{K}+\bar {\epsilon }_{\mu }}(v_2)
=\sum _{\mu ',\nu '}
\bar {W}(\widehat{K}|v_{1}-v_{2})^{\mu '\nu '}_{\mu\nu}
\tilde {\varphi }^{(i)}_{\mu ',\widehat{K}+\bar {\epsilon }_{\nu '}}(v_1)
\tilde {\varphi }^{(j)}_{\nu ',\widehat{K}}(v_2),\nonumber\\
\end{eqnarray}
\begin{eqnarray}
&&\sum _{k,l}\bar {R'}^{(1)}(v_1-v_2)^{ij}_{kl}
\bar {\varphi '}^{(k)}_{\widehat{L}-\bar {\epsilon }_{\nu },\mu }(v_1)
\bar {\varphi '}^{(l)}_{\widehat{L},\nu }(v_2)
=\sum _{\mu ',\nu '}
\bar {W'}(\widehat{L}|v_{1}-v_{2})^{\nu'\mu'}_{\mu\nu}
\bar {\varphi '}^{(i)}_{\widehat{L},\nu '}(v_1)
\bar {\varphi '}^{(j)}_{\widehat{L}-\bar {\epsilon }_{\nu '},\mu '}(v_2).\nonumber\\
\end{eqnarray}
Introduce the modular transformation
\begin{eqnarray}
\theta \left[
\begin{array}{c}
{1\over 2}+a\\
{1\over 2}+b\end{array}\right]({z\over \tau},-{1\over \tau})
=\theta \left[
\begin{array}{c}
{1\over 2}+b\\ {1\over 2}-a\end{array}\right] (z,\tau )
exp\pi i\{ \frac {z^2}{\tau}+a-b+2ab\}\times const. ,
\end{eqnarray}
\noindent where the const. only  depends on $\tau$.
Therefore ,we can derive the following relations for $Z_n$ symmetry 
R-matrices $\bar{R}^{(1)}(v)$ and $\bar{R}'^{(1)}(v)$
\begin{eqnarray}
& &(M\otimes M)\bar{R}^{(1)}(v)(M^{-1}\otimes M^{-1})=x^{\frac{2v(1-n)}{nr}}
P\bar{R}(vw,rw)P,\\
& &(M\otimes M)\bar{R}'^{(1)}(v)(M^{-1}\otimes M^{-1})=
x^{\frac{2v(1-n)}{n(r-1)}}P\bar{R}(vw,(r-1)w)P,
\end{eqnarray}
\noindent where $P$ is the permutation operator acting on the tensor space 
$V\otimes V$ as :$P(e_{i}\otimes e_{j})=e_{j}\otimes e_{i}$,and the matrix 
$(M)_{lk}=\omega^{-lk}=exp\{\frac{-2i\pi}{n}lk\}$ which have the following 
properties 
\begin{eqnarray*}
MgM^{-1}=h^{-1}\ \ ,\ \ MhM^{-1}=g.
\end{eqnarray*}

\subsection{Bosonization for $Z_n$ Belavin model}
Based on the bosonization for $A^{(1)}_{n}$ face model and the face-vertex 
correspondence , we will construct the bosonization of two type vertex 
operatos for $Z_n$ Belavin model .

Firstly, define
\begin{eqnarray}
& &\Phi_{j}(v)=\sum_{i=1}^{n}\sum_{\mu=1}^{n}M_{ji}\phi_{\mu}(v)
\tilde {\varphi}^{(i)}_{\mu ,\widehat{K}}(-v),\\
& &\Psi_{j}(v)=\sum_{i=1}^{n}\sum_{\mu=1}^{n}M_{ji}\phi '_{\mu}(v)
\bar {\varphi}^{(i)}_{-\widehat{L},\mu}(-v),
\end{eqnarray}
\noindent which satisfy the commutation relations on the space 
$\sum_{l,k\in P}\oplus F_{l,k}$
\begin{eqnarray}
\Phi _j(v_2)\Phi _i(v_1)&=&\sum_{ij}
R_{lk}^{ij}(v_1-v_2)\Phi _l(v_1)\Phi _k(v_2),
\\
\Psi _i(v_1)\Psi _j(v_2)&=&\sum_{lk}\Psi _k(v_2)\Psi _l(v_1)
{R^*}_{lk}^{ij}(v_1-v_2)\Delta _n^{-1}(v_1-v_2),
\\
\Phi _i(v_1)\Psi _j(v_2)&=&\tau (v_1-v_2)
\Psi _j(v_2)\Phi _i(v_1).
\end{eqnarray}
Based on the anti-symmetric fusion for $Z_n$ symmetric R-matrix [10,15],
we can define  
\begin{eqnarray*}
\Psi^{*}_{j}(v)=\Psi_{(1,...,j-1,j+1,...,n)}(v)\equiv 
\sum_{\sigma \in S_{n-1}}sign(\sigma )\Psi_{\sigma(1)}(v+n-1)
\Psi_{\sigma(2)}(v+n-2)....\Psi_{\sigma(n)}(v+1),
\end{eqnarray*}
and derive the following relations
\begin{eqnarray}
\Phi _j(v_2)\Phi _i(v_1)&=&\sum_{lk}R_{lk}^{ij}(v_1-v_2)
\Phi _l(v_1)\Phi _k(v_2),
\\
\Psi _i^*(v_1)\Psi _j^*(v_2)&=&\sum_{lk}\Psi _k^*(v_2)\Psi _l^*(v_1)
{R^*}^{lk}_{ij}(v_1-v_2)\Delta _n^{-1}(v_1-v_2),
\\
\Phi _i(v_1)\Psi _j^*(v_2)&=&\tau ^{-1}(v_1-v_2)
\Psi _j^*(v_2)\Phi _i(v_1).
\end{eqnarray}
Therefore, we obtain the bosonic realization of vertex operators 
$\Phi_{i}(v)$ and $\Psi^{*}_{i}(v)$ defined in section 2.Morever,
we obtain the bosonic realization for $A_{q,p}(\widehat {sl_n})$ 
algebra at level one through Miki construction Eq.(16).Thus this 
algebra is self-consistent.

{\bf Remark :}For the case n=2 ($\Delta_{2}=-1$),the bosonic operators 
$\Phi_{i}(v)$ and $\Psi^{*}_{i}(v)$ become the type I and type II vertex 
operators of $A_{q,p}(\widehat {sl_2})$ at level one proposed by Foda et 
al [3].

\section*{Discussions}
For n=2 ,the algebra $A_{q,p}(\widehat {sl_n})$ reduces to the original 
one $A_{q,p}(\widehat {sl_2})$ algebra which was first proposed by 
Foda et al[3].The quantum affine algebra $U_{q}(\widehat {sl_2})$ 
is a degeneration algebra of $A_{q,p}(\widehat {sl_2})$ with $p=0$ [3].
Unfortunately,due to the nontrivial scalar factor $\Delta_{n}(v)$ when 
$2<n$ ,the relation between the elliptic algebra $A_{q,p}
(\widehat {sl_n})$ and $U_{q}(\widehat {sl_{n}})$ is still an open problem.

As discussed in [8], vertex operators $\phi_{\mu}(v)$ in $A^{(1)}_{n-1}$ face 
model was the q-analog of the chiral primary fields of q-deformed 
W-algebras[14].However,the vertex operators $\Phi_{i}(v)$ and 
$\Psi^{*}_{j}(v)$ in $Z_n$ Belavin model are reconstructed by the vertex
 operators of $A^{(1)}_{n-1}$ face model through the face-vertex  
correspondence  relations.Moreover,the elliptic algebra 
$A_{q,p}(\widehat {sl_n})$ at level one can be constructed by Miki 
construction .So, there exists some relations between the algebra 
$A_{q,p}(\widehat {sl_n})$ and q-deformed W-algebras.

Besides the degenerate algebra $A_{q,0}(\widehat {sl_n})$ ,there exists 
another degenerate algebra $A_{\hbar ,\eta}(\widehat {sl_n})$ [16],which 
can be obtained by the scaling limit ($v=\frac{u}{\hbar},q=x,
p=x^{\eta},x\longrightarrow 1)$ of the elliptic algebra 
$A_{q,p}(\widehat {sl_n})$. The resulted algebra 
$A_{\hbar ,\eta}(\widehat {sl_n})$ is some deformation of Yangian double 
$DY(\widehat {sl_n})$ [16,17].In  the scaling limit case the R-matrix 
entering the $\Psi^{*}\Psi^{*}$ commutation relation would be 
interpreted as the S-matrix for soliton of affine Toda theroy 
(In the case of n=2 ,it is related to sine-Gordon theory [17]).  
Moreover, the $\hbar$-deformed W-algebra [18] would be obtained 
by the quantum Hamiltonian reduction from the degenerate algebra 
$A_{\hbar ,\eta}(\widehat {sl_n})$.

In our formulation, the algebra $A_{q,p}(\widehat {sl_{n}})$ is formulated  
in the framework of the $``RLL"$ approach in terms of the L-operator.
It would be of great importance to find an analogic Drinfeld currents 
for the algebra $A_{q,p}(\widehat {sl_n})$.For a special case 
$A_{q,p}(\widehat {sl_2})$ with the scaling limit,the Drinfeld currents for  
the algebra $A_{\hbar ,\eta}(\widehat {sl_2})$ 
was found to be the Guass coordinates of the L-operator[16].We expect 
that the same relation  would be existed for the elliptic algebra 
$A_{q,p}(\widehat {sl_n})$.

\section*{Appendix}
\subsection*{A. The normal order relation for basic operator}
We list the relations as three series:

{\bf Type I}:
\begin{eqnarray}
\eta _1(v_1)\eta _1(v_2)
&=&x^{\frac {2(r-1)(n-1)v_1}{nr}}g_1(v_2-v_1):\eta _1(v_1)\eta _1(v_2):
\nonumber \\
\eta _j(v_1)\xi _j(v_2)
&=&x^{\frac {2(1-r)}{r}v_1}s(v_2-v_1):\eta _j(v_1)\xi _j(v_2):
\nonumber \\
\xi _j(v_2)\eta _j(v_1)
&=&x^{\frac {2(1-r)}{r}v_2}s(v_1-v_2):\xi _j(v_2)\eta _j(v_1):
\nonumber \\
\xi _j(v_1)\xi _{j+1}(v_2)
&=&x^{\frac {2(1-r)}{r}v_1}s(v_2-v_1):\xi _j(v_1)\xi _{j+1}(v_2):
\nonumber \\
\xi _{j+1}(v_2)\xi _j(v_1)
&=&x^{\frac {2(1-r)}{r}v_2}s(v_1-v_2):\xi _j(v_1)\xi _{j+1}(v_2):
\nonumber \\
\xi _j(v_1)\xi _j(v_2)
&=&x^{\frac {4(r-1)}{r}v_1}t(v_2-v_1):\xi _j(v_1)\xi _j(v_2):
\nonumber \\
\xi _j(v_1)\xi _l(v_2)
&=&:\xi _j(v_1)\xi _l(v_2):~~~{\rm if}~|l-j|>1,
\nonumber \\
\eta _j(v_1)\xi _l(v_2)
&=&:\eta _j(v_1)\xi _l(v_2):~~~{\rm if}~l\not= j,
\nonumber \\
\end{eqnarray}

{\bf Type II}:
\begin{eqnarray}
\eta '_1(v_1)\eta '_1(v_2)
&=&x^{\frac {2r(n-1)v_1}{(r-1)n}}g'_1(v_2-v_1):\eta '_1(v_1)\eta '_1(v_2):
\nonumber \\
\xi '_j(v_1)\eta '_j(v_2)
&=&x^{\frac {-2r}{r-1}v_1}s'(v_2-v_1):\xi '_j(v_1)\eta '_j(v_2):
\nonumber \\
\eta '_j(v_2)\xi '_j(v_1)
&=&x^{\frac {-2r}{r-1}v_2}s'(v_1-v_2):\eta '_j(v_2)\xi '_j(v_1):
\nonumber \\
\xi '_j(v_1)\xi '_{j+1}(v_2)
&=&x^{\frac {-2r}{r-1}v_1}s'(v_2-v_1):\xi '_j(v_1)\xi '_{j+1}(v_2):
\nonumber \\
\xi '_{j+1}(v_2)\xi '_j(v_1)
&=&x^{\frac {-2r}{r-1}v_2}s'(v_1-v_2):\xi '_j(v_1)\xi '_{j+1}(v_2):
\nonumber \\
\xi '_j(v_1)\xi '_j(v_2)
&=&x^{\frac {4r}{r-1}v_1}t'(v_2-v_1):\xi '_j(v_1)\xi '_j(v_2):
\nonumber \\
\xi '_j(v_1)\xi '_l(v_2)
&=&:\xi '_j(v_1)\xi '_l(v_2):~~~{\rm if}~|l-j|>1,
\nonumber \\
\eta '_j(v_1)\xi '_l(v_2)
&=&:\eta '_j(v_1)\xi '_l(v_2):~~~{\rm if}~l\not= j,
\nonumber \\
\end{eqnarray}

{\bf Type I with type II}:
\begin{eqnarray}
\eta _j(v_1)\xi '_j(v_2)
&=&
(x^{2v_1}-x^{2v_2}):\xi '_j(v_2)\eta _j(v_1):
\nonumber \\
\xi '_j(v_2)\eta _j(v_1)
&=&
(x^{2v_2}-x^{2v_1}):\eta _j(v_1)\xi '_j(v_2):
\nonumber \\
\xi _j(v_1)\eta '_j(v_2)
&=&(x^{2v_1}-x^{2v_2}):\xi _j(v_1)\eta '_j(v_2):
\nonumber \\
\eta '_j(v_2)\xi _j(v_1)
&=&(x^{2v_2}-x^{2v_1})
:\eta '_j(v_2)\xi _j(v_1):
\nonumber \\
\eta _l(v_1)\xi '_j(v_2)&=&:\eta _l(v_1)\xi '_j(v_2):
=\xi '_j(v_2)\eta _l(v_1)~~~{\rm if}~l\not= j,
\nonumber \\
\eta '_l(v_1)\xi _j(v_2)&=&:\eta '_l(v_1)\xi _j(v_2):
=\xi _j(v_2)\eta '_l(v_1)~~~{\rm if}~l\not= j,
\nonumber \\
\xi _j(v_1)\xi '_{j+1}(v_2)
&=&
(x^{2v_1}-x^{2v_2}):\xi _j(v_1)\xi '_{j+1}(v_2):
\nonumber \\
\xi '_{j+1}(v_2)\xi _j(v_1)
&=&(x^{2v_2}-x^{2v_1}):\xi '_{j+1}(v_2)\xi _j(v_1):
\nonumber \\
\xi _j(v_1)\xi '_j(v_2)
&=&x^{-4v_1}
\frac {1}{(1-xx^{2(v_2-v_1)})(1-x^{-1}x^{2(v_2-v_1)})}
:\xi _j(v_1)\xi '_j(v_2):
\nonumber \\
\xi '_j(v_2)\xi _j(v_1)
&=&x^{-4v_2}
\frac {1}{(1-xx^{2(v_1-v_2)})(1-x^{-1}x^{2(v_1-v_2)})}
:\xi '_j(v_2)\xi _j(v_1):
\nonumber \\
\xi _j(v_1)\xi '_l(v_2)&=&:\xi _j(v_1)\xi '_l(v_2):~~~{\rm if}~|j-l|>1,
\nonumber \\
\xi '_l(v_2)\xi _j(v_1)&=&:\xi '_l(v_2)\xi _j(v_1):~~~{\rm if}~|j-l|>1,
\nonumber \\
\eta _1(v_1)\eta '_j(v_2)&=&
x^{2v_1\frac {j-n}{n}}
\frac {(x^{2n-j}x^{2(v_2-v_1)};x^{2n})}
{(x^jx^{2(v_2-v_1)};x^{2n})}
:\eta _1(v_1)\eta '_j(v_2):
\nonumber \\
\eta '_j(v_2)\eta _1(v_1)&=&
x^{2v_2\frac {j-n}{n}}
\frac {(x^{2n-j}x^{2(v_1-v_2)};x^{2n})}
{(x^jx^{2(v_1-v_2)};x^{2n})}
:\eta '_j(v_2)\eta _1(v_1):
\end{eqnarray}
where 
\begin{eqnarray}
& &g_1(v)=\frac {\{ x^2x^{2v}\} \{ x^{2n+2r-2}x^{2v}\} }
{\{ x^{2n}x^{2v}\} \{ x^{2r}x^{2v}\} },\ \ 
s(v)=\frac {(x^{2r-1}x^{2v};x^{2r})}
{(xx^{2v};x^{2r})},
\nonumber \\
& &t(v)=(1-x^{2v})
\frac {(x^2x^{2v};x^{2r})}
{(x^{2r-2}x^{2v};x^{2r})},
\nonumber\\ 
& &g'_1(v)=\frac {\{ x^{2v}\} '\{ x^{2n+2r-2}x^{2v}\} '}
{\{ x^{2r}x^{2v}\} '\{ x^{2n-2}x^{2v}\} '},
~~\{ z\} '=(z;x^{2r-2}, x^{2n})
\nonumber \\
& &s'(v)=\frac {(x^{2r-1}x^{2v};x^{2r-2})}
{(x^{-1}x^{2v};x^{2r-2})},
\ \ t'(v)=(1-x^{2v})
\frac {(x^{-2}x^{2v};x^{2r-2})}
{(x^{2r}x^{2v};x^{2r-2})}.
\end{eqnarray}
\subsection*{B. Proof of the commutation relations Eq.(43)-Eq.(45)}
Notice that the commutation relation for zero mode operators,we have 
\begin{eqnarray*}
& &\phi_{\nu}(v_{1})\tilde{\varphi}_{\mu,\widehat{K}}^{(k)}(v_{2})
=\tilde{\varphi}
_{\mu,\widehat{K}-\overline{\epsilon}_{\nu}}^{(k)}(v_{2})
\phi_{\nu}(v_{1})\ \ ,\ \ 
\phi'_{\nu}(v_{1})\tilde{\varphi}_{\mu,\widehat{K}}^{(k)}(v_{2})
=\tilde{\varphi}_{\mu,\widehat{K}}^{(k)}(v_{2})\phi'_{\nu}(v_{1}),\\
& &\phi_{\nu}(v_{1})\bar {\varphi '}^{(k)}_{\widehat{L},\mu }(v_2)
=\bar {\varphi '}^{(k)}_{\widehat{L},\mu }(v_2)\phi_{\nu}(v_{1})\ \ ,\ \ 
\phi'_{\nu}(v_{1})\bar {\varphi '}^{(k)}_{\widehat{L},\mu }(v_2)
=\bar {\varphi '}^{(k)}_{\widehat{L}-\overline{\epsilon}_{\nu},\mu }(v_2)
\phi'_{\nu}(v_{1}).
\end{eqnarray*}
Define
\begin{eqnarray*}
& &Z_{j}(v)=\sum_{\mu=1}^{n}\phi_{\mu}(v)
\tilde {\varphi}^{(i)}_{\mu ,\widehat{K}}(-v)\ \ ,\ \ 
Z'_{j}(v)=\sum_{\mu=1}^{n}\phi '_{\mu}(v)
\bar {\varphi}^{(i)}_{-\widehat{L},\mu}(-v).
\end{eqnarray*}
Notice that 
\begin{eqnarray*}
\bar{W}(\widehat{\pi}|v)^{\nu'\mu'}_{\nu\mu}|_{F_{l,k}}=
\bar{W}(\widehat{K}|v)^{\nu'\mu'}_{\nu\mu}|_{F_{l,k}}
\end{eqnarray*}
using the exchange relation of $\phi_{\mu}(v)$ in Eq.(32) and the face-vertex 
correspondence by $\tilde{\varphi}^{(k)}_{\mu ,\widehat{K}}(v)$ in Eq.(36), 
we have  
\begin{eqnarray*}
Z_{j} (v_{2})Z_{i} (v_{1})=\sum_{l,k}r_{1}(v_{2}-v_{1})(P\bar{R}^{(1)}
(v_{1}-v_{2})P)^{ij}_{lk}Z_{l}(v_{1})Z_{k}(v_{2}).
\end{eqnarray*}
Notice the properties of $Z_n$ symmetry R-matrix under the modular 
transformation Eq.(39) ,we obtain  
\begin{eqnarray*}
Z_{j} (v_{2})Z_{i} (v_{1})=\sum_{l,k}(M^{-1}\otimes M^{-1}R(v_{1}-v_{2})
M\otimes M)^{ij}_{lk}Z_{l}(v_{1})Z_{k}(v_{2})
\end{eqnarray*}
Due to $\Phi_{j}(v)=\sum_{i=1}^{n}M_{ji}Z_{i}(v)$ , one can obtain Eq.(43).
Using the same method and notice that
\begin{eqnarray*}
& &\bar {W'}(\widehat{L}|v)^{\mu\nu}_{\nu'\mu'}=\bar{W'}(-\widehat{L}|v)
^{\nu\mu}_{\mu'\nu'}\ \ ,\ \ 
\frac{r'_{1}(v)}{r_{1}(-v)|_{r\longrightarrow r-1}}=\Delta_{n}(v),\\
& &\bar{W}'(\widehat{\pi}|v)^{\nu'\mu'}_{\nu\mu}|_{F_{l,k}}=
\bar{W}'(\widehat{L}|v)^{\nu'\mu'}_{\nu\mu}|_{F_{l,k}}
\end{eqnarray*}
we can obtain Eq.(44) and Eq.(45).

\end{document}